\documentclass[twocolumn]{autart}
\usepackage{graphicx}          
\usepackage{lineno,hyperref}
\usepackage{bm,cite}
\usepackage{algorithm}
\usepackage{multirow}
\usepackage{multicol}
\usepackage{booktabs}
\modulolinenumbers[5]
\usepackage{amsmath,amsfonts,amssymb}
\begin{document}

\begin{frontmatter}
\title{Minimum Error Entropy Kalman Filter}
\thanks[footnoteinfo]{This paper was not presented at any IFAC
meeting. Corresponding author Badong~Chen. Tel. 86-29-82668802 ext.8009. Fax. 86-29-82668672.}

\author[Paestum]{Badong~Chen}\ead{chenbd@mail.xjtu.edu.cn},    
\author[Paestum]{Lujuan~Dang}\ead{danglj@stu.xjtu.edu.cn},               
\author[Rome]{Yuantao~Gu}\ead{gyt@tsinghua.edu.cn},
\author[Paestum]{Nanning~Zheng}\ead{nnzheng@mail.xjtu.edu.cn},   \author[Paestum,Baiae]{Jos\'{e} C. Pr\'{i}ncipe}\ead{principe@cnel.ufl.edu}

\address[Paestum]{Institute of Artificial Intelligence and Robotics, Xi'an Jiaotong University, Xi'an, 710049, China}
\address[Rome]{Beijing National Research Center for Information
Science and Technology (BNRist) and Department of Electronic Engineering, Tsinghua University, Beijing, 100084, China}
\address[Baiae]{Department of Electrical and Computer Engineering, University of Florida, Gainesville, FL, 32611, USA}


\begin{abstract}
To date most linear and nonlinear Kalman filters (KFs) have been developed under the Gaussian assumption and the well-known minimum mean square error (MMSE) criterion. In order to improve the robustness with respect to impulsive (or heavy-tailed) non-Gaussian noises, the maximum correntropy criterion (MCC) has recently been used to replace the MMSE criterion in developing several robust Kalman-type filters. To deal with more complicated non-Gaussian noises such as noises from multimodal distributions, in the present paper we develop a new Kalman-type filter, called minimum error entropy Kalman filter (MEE-KF), by using the minimum error entropy (MEE) criterion instead of the MMSE or MCC. Similar to the MCC based KFs, the proposed filter is also an online algorithm with recursive process, in which the propagation equations are used to give prior estimates of the state and covariance matrix, and a fixed-point algorithm is used to update the posterior estimates. In addition, the minimum error entropy extended Kalman filter (MEE-EKF) is also developed for performance improvement in the nonlinear situations. The high accuracy and strong robustness of MEE-KF and MEE-EKF are confirmed by experimental results.
\end{abstract}

\end{frontmatter}

\section{Introduction}
Kalman filtering is a powerful technology for estimating the states of a dynamic system, which finds applications in many areas including navigation, guidance, data integration, pattern recognition, tracking and control systems \cite{KF1,KF,KFT,KF2,KF3}. The original Kalman filter (KF) was derived for a linear state space model with Gaussian assumption \cite{KF44, KF}. To cope with nonlinear estimation problems, a variety of nonlinear extensions of the original Kalman filter have been proposed in the literature, including extended Kalman filter (EKF) \cite{EKF,EKF11}, unscented Kalman filter (UKF) \cite{UKF}, cubature Kalman filter (CKF) \cite{CKF} and many others. However, most of these Kalman filters are developed based on the popular minimum mean square error (MMSE) criterion and will face performance degradation in case of complicated noises, since in general MMSE is not a good choice for the estimation in non-Gaussian noises.

In recent years, to solve the performance degradation problem in heavy-tailed (or impulsive) non-Gaussian noises, some robust Kalman filters have been developed by using certain non-MMSE criterion as the optimality criterion \cite{JCITL,CBDITL}. Particularly, the maximum correntropy criterion (MCC) \cite{C,MCC111} in information theoretic learning (ITL) \cite{JCITL,CBDITL} has been successfully applied in Kalman filtering to improve the robustness against impulsive noises. Typical examples include the maximum correntropy based Kalman filters \cite{KFMCchen,MCCKF2,MCCKF3,MCCKF4,MCCKF5,MCCKF55,MCCKF6,MCCKF7,MCCKF8,MCCKF9}, maximum correntropy based extended Kalman filters \cite{MCEKF,MCEKF1,MCEKF2}, maximum correntropy based unscented Kalman filters \cite{MCUKF,MCUKF1,MCUKF2}, maximum correntropy based square-root cubature Kalman filters \cite{MCSCKF,MCSCKF1} and so on. Since correntropy is a local similarity measure and insensitive to large errors, these MCC based filters are little influenced by large outliers \cite{C,MCCMAP}.

The MCC is a nice choice for dealing with heavy-tailed non-Gaussian noises, but its performance may not be good when facing more complicated non-Gaussian noises, such as noises from multimodal distributions. The minimum error entropy (MEE) criterion \cite{MEE,CMEE} is another important learning criterion in ITL, which has been successfully applied in robust regression, classification, system identification and adaptive filtering \cite{MEE,CMEE,GIPC,EEC,MECQ}. Numerous experimental results show that MEE can outperform MCC in many situations although its computational complexity is a little higher \cite{SIP,KMEE}. In addition, the superior performance and robustness of MEE have been proved in \cite{RobustMEE}. The goal of this work is to develop a new Kalman-type filter, called minimum error entropy Kalman filter (MEE-KF), by using the MEE as the optimality criterion. The proposed filter uses the propagation equations to obtain the prior estimates of the state and covariance matrix, and a fixed-point algorithm to update the posterior estimates and covariance matrix, recursively and online. To further improve the performance in the nonlinear situations, the MEE criterion is also incorporated into EKF, resulting in minimum error entropy extended Kalman filter (MEE-EKF).

The rest of the paper is organized as follows. In section II, we briefly review the KF algorithm and MEE criterion. In section III, we develop the  MEE-KF algorithm. Sections IV and V provide the computational complexity and convergence analysis, respectively.  In section VI, the MEE-EKF is developed. The experimental results are presented in section VII and finally, the conclusion is given in section VIII.
\section{Background}
\subsection{Kalman Filter}
Consider a linear dynamic system with unknown state vector $\textbf{x}(k)\in\mathbb{R}^{n\times1}$ and available measurement vector $\textbf{y}(k)\in\mathbb{R}^{m\times1}$. To estimate the state $\textbf{x}(k)$, Kalman filter (KF) assumes a state space model described by the following state and measurement equations:
\begin{align}
&\textbf{x}(k){\rm{ = }}\textbf{F}\textbf{x}(k - 1){\rm{ + }}\textbf{q}(k - 1)\label{state}\\
&\textbf{y}(k){\rm{ = }}\textbf{H}\textbf{x}(k) + \textbf{r}(k),\label{measurement}
\end{align}
where $\textbf{F}\in\mathbb{R}^{n\times n}$ and $\textbf{H}\in\mathbb{R}^{m\times n}$ are the state-transition matrix and measurement matrix, respectively. Here, the process noise $\textbf{q}(k - 1)\in\mathbb{R}^{n\times1}$ and measurement noise $ \textbf{r}(k)\in\mathbb{R}^{m\times1}$ are mutually independent, and satisfy
\begin{align}
&E[\textbf{q}(k - 1)]=0;  E[\textbf{r}(k)]=0\\
&E[\textbf{q}(k - 1)\textbf{q}^{\textrm{T}}(k - 1)]=\textbf{Q}(k - 1)\\
&E[\textbf{r}(k )\textbf{r}^{\textrm{T}}(k )]=\textbf{R}(k),
\end{align}
where  $\textbf{Q}(k - 1)$ and $\textbf{R}(k)$ are the covariance matrices of $\textbf{q}(k - 1)$ and $\textbf{r}(k)$, respectively. In general, the KF includes two steps:

(1) Predict: The \emph{a-priori} estimate $\hat{\textbf{x}}(k|k - 1)$ and the corresponding error covariance matrix $\textbf{P}(k|k - 1)$ are calculated by
\begin{align}
&\hat{\textbf{x}}(k|k - 1){\rm{ = }}\textbf{F} {\hat {\textbf{x}}}(k - 1)\label{Predict1}\\
&\textbf{P}(k|k - 1){\rm{ = }}\textbf{F}\textbf{P}(k - 1){\textbf{F}^{\rm{T}}}{\rm{ + }}\textbf{Q}(k - 1).\label{Predict2}
\end{align}

(2) Update: The \emph{a-posteriori} estimate ${\hat {\textbf{x}}}(k)$ and the corresponding error covariance matrix $\textbf{P}(k)$ are obtained by
\begin{align}
&{\hat {\textbf{x}}}(k) = {\hat {\textbf{x}}}(k|k - 1) + {\textbf{K}}(k)[\textbf{y}(k) - \textbf{H}{\hat {\textbf{x}}}(k|k - 1)]\label{update1}\\
&\textbf{K}(k){\rm{ = }}\textbf{P}({{k|}}k - 1){\textbf{H}^\textrm{T}}[\textbf{H}\textbf{P}({{k|}}k - 1){\textbf{H}^\textrm{T}} + \textbf{R}(k)]^{-1}\label{update3}\\
&\textbf{P}(k){\rm{ = [}}\textbf{I} - {\textbf{K}}(k)\textbf{H}]\textbf{P}(k|k - 1){{\rm{[}}\textbf{I} - {\textbf{K}}(k)\textbf{H}]^\textrm{T}}\nonumber\\
&\,\,\,\,\,\,\,\,\,\,\,\,\,\,\,+ {\textbf{K}}(k)\textbf{R}(k){\textbf{K}^{\textrm{T}}}(k)\label{update2},
\end{align}
where $\textbf{K}(k)$ is the Kalman filter gain.
\subsection{Minimum Error Entropy Criterion}
Different from the MMSE \cite{MEE} and MCC \cite{CMEE}, the MEE aims to minimize the information contained in the error. In MEE, the error information $e$ can be measured by the Renyi's entropy:
\begin{align}\label{Renyientropy}
{{{H}}_\alpha }(e) =  \frac{1}{{1 - \alpha }}\log {V_\alpha }(e),
\end{align}
where $\alpha (\alpha  \ne {\rm{1}}, \alpha  > 0 )$ is the order of Renyi's entropy, and ${V_\alpha }(e)$ denotes the information potential defined by
\begin{align}\label{informationpotential}
{V_\alpha }(e) =\int { p^{\alpha} (x )} dx= {E}[ p^{\alpha  - 1}(e)],
\end{align}
where $p(.)$ is the probability density function (PDF) of error $e$ and
${E}[\cdot]$ denotes the expectation operator. In practical applications, the PDF $p(x)$ can be estimated by Parzen's window approach \cite{JCITL}:
\begin{align}\label{p(e)}
 \hat p(x) = \frac{1}{N}\sum\limits_{i = 1}^N {{G_\sigma }(x - e_i)},
\end{align}
where ${{G_\sigma }(x)}=\frac{1}{\sqrt{2 \pi} \sigma}\exp \left(- (x)^2\over {2\sigma^2} \right)$ denotes the Gaussian kernel with kernel size $\sigma$; $\{e_i\}_{i=1}^N$ are $N$ error samples.
Combining (\ref{informationpotential}) and (\ref{p(e)}), one can obtain an estimate of the second order ($\alpha=2$) information potential $ V_2(e)$:
\begin{align}\label{Vz}
\hat V_2(e) =\frac{1}{N}\sum\limits_{i = 1}^N \hat p(e_i)=\frac{1}{{{N^2}}}\sum\limits_{i = 1}^N {\sum\limits_{j = 1}^N {{G_\sigma }({e_i} - {e_j})}}.
\end{align}
Since the negative logarithmic function $-\textrm{log}$ is monotonically decreasing, minimizing the error entropy ${{{H}}_2 }(e)$ means maximizing the information potential ${\hat V_2 }(e)$.
\section{Minimum Error Entropy Kalman Filter}
\subsection{Augmented Model}
First, we denote the state prediction error as
\begin{align}\label{18}
{\bm{\epsilon}}(k|k - 1)= \textbf{x}(k)-{\hat {\textbf{x}}}(k|k - 1).
\end{align}
Combining the above state prediction error with the measurement equation (\ref{measurement}), one can obtain an augmented model
\begin{align}\label{augmentedmodel}
\begin{array}{c}
\left[ \begin{array}{l}
{\hat {\textbf{x}}}(k|k - 1)\\
\textbf{y}(k)
\end{array} \right] = \left[ \begin{array}{l}
\textbf{I}_{n}\\
\textbf{H}
\end{array} \right]\textbf{x}(k) +\mu(k),
\end{array}
\end{align}
where $\textbf{I}_{n}$ denotes a $n\times n$ identity matrix, and
\begin{align}\label{20}
\mu(k)=
\left[ \begin{array}{l}
 - {{\bm{\epsilon}}}(k|k - 1)\\
\textbf{r}(k)
\end{array} \right]
\end{align}
is the augmented noise vector comprising of the state and measurement errors. Assuming that the covariance matrix of the augmented noise ${\rm{E}}\left[ {\mu (k)\mu  {{(k)}^\textrm{T}}} \right]$ is positive definite, we have
\begin{align}\label{211}
\begin{array}{l}
{\rm{E}}\left[ {\mu (k)\mu  {{(k)}^\textrm{T}}} \right]
 = {\bm{\Theta}}(k){\bm{\Theta}}{(k)^\textrm{T}}\\
 = \left[ \begin{array}{l}
\begin{array}{*{20}{c}}
{{{\bm{\Theta}}_p}(k|k - 1){{\bm{\Theta}}_p}{{(k|k - 1)}^\textrm{T}}\;\;\;\;}&\textbf{0}
\end{array}\\
\begin{array}{*{20}{c}}
\textbf{0}&{\;\;\;\;\;\;\;\;\;\;\;\;\;\;\;\;\;\;\;\;\;\;\;\;\;\;\;\;\;\;\;\;\;\;\;{{\bm{\Theta}}_r}(k){{\bm{\Theta}}_r}{{(k)}^\textrm{T}}}
\end{array}
\end{array} \right],\\
\end{array}
\end{align}
where ${\bm{\Theta}}(k)$, ${{\bm{\Theta}}_p}{(k|k - 1)}$ and ${{\bm{\Theta}}_r}(k)$ are obtained by the Cholesky decomposition of ${\rm{E}}\left[ {\mu (k)\mu  {{(k)}^\textrm{T}}} \right]$, $\textbf{P}(k|k - 1)$ and $\textbf{R}(k)$, respectively.
Multiplying both sides of (\ref{augmentedmodel}) by ${{\bm{\Theta}}^{{\rm{ - }}1}}(k)$ gives
\begin{align}\label{equation23}
\textbf{d}(k) &= \textbf{W}(k)\textbf{x}(k) + \textbf{e}(k),
\end{align}
where
\begin{align}
&\textbf{d}(k){\rm{ = }}{{\bm{\Theta}}^{{\rm{ - }}1}}(k)\left[ \begin{array}{l}
{\hat {\textbf{x}}}(k|k - 1)\\
\textbf{y}(k)
\end{array} \right]\label{21}\\
&\textbf{W}(k) = {{\bm{\Theta}}^{{\rm{ - }}1}}(k)\left[ \begin{array}{l}
\textbf{I}_{n}\\
\textbf{H}
\end{array} \right]\label{22}\\
&\textbf{e}(k){\rm{ = }}{{\bm{\Theta}}^{{\rm{ - }}1}}(k)\left[ \begin{array}{l}
 - (\textbf{x}(k) - {\hat {\textbf{x}}}(k|k - 1))\\
\textbf{r}(k)
\end{array} \right]\label{27},
\end{align}
with
$\textbf{d}(k) = {{\rm{[}}{{d}_1}(k){\rm{,}}{{d}_2}(k),...,{{d}_L}(k){\rm{]}}^\textrm{T}}$,
$\textbf{W}(k) =
[{\textbf{w}_1}(k),{\textbf{w}_2}(k),\\...,{\textbf{w}_L}(k)]^\textrm{T}$,
$\textbf{e}(k){\rm{ = [}}{{e}_1}(k){\rm{,}}{{e}_2}(k),...,{{e}_L}(k){{\rm{]}}^\textrm{T}}$ and $L=m+n$.

\subsection{Derivation of MEE-KF}
Based on (\ref{Vz}), the cost function of MEE-KF is given by
\begin{align}\label{MEEcostfunction}
{J_{{L}}}{\rm{(}}\textbf{x}(k)) = \frac{1}{{{L^{\rm{2}}}}}\sum\limits_{i = 1}^L {\sum\limits_{j = 1}^L {{G_\sigma }(e{}_j(k) -  e{}_i(k))}}.
\end{align}
Then, the optimal solution to ${\hat {\textbf{x}}}(k)$ is achieved by maximizing the cost function (\ref{MEEcostfunction}), that is
\begin{align}
&{\hat {\textbf{x}}}(k){\rm{ = arg }}\mathop {{\rm{max}}}\limits_{\textbf{x}(k)}  {J_L}(\textbf{x}(k))\nonumber\\
 &= {\rm{arg }}\mathop {{\rm{max}}}\limits_{\textbf{x}(k)}
\frac{1}{{{L^{\rm{2}}}}}\sum\limits_{i = 1}^L {\sum\limits_{j = 1}^L {{G_\sigma }(e{}_j(k) -  e{}_i(k))}}.
\end{align}
Setting the gradient of the cost function ${J_{{L}}}{\rm{(}}\textbf{x}(k))$ regarding $\textbf{x}(k)$ to zero, we have
\begin{align}\label{error111}
&\frac{\partial }{{\partial \textbf{x}(k)}}{J_L}{\rm{(}}\textbf{x}(k){\rm{)}}  \nonumber\\ &=\frac{1}{{{L^2\sigma^2}}}\sum\limits_{i = 1}^L {\sum\limits_{j = 1}^L {\left( \begin{array}{l}
[e{}_j(k) -  e{}_i(k)]{G_\sigma }(e{}_j(k) - e{}_i(k))\\
{[{\textbf{w}_j}(k) -  {\textbf{w}_i}(k)]}
\end{array} \right)} } \nonumber\\
 &= \Gamma_1- \Gamma_2- \Gamma_3+ \Gamma_4=  2\Gamma_1-2\Gamma_3 \nonumber\\
 &= \frac{2}{{{L^2\sigma^2}}}\textbf{W}{(k)^\textrm{T}}{\bm{\Psi}} (k)\textbf{e}(k) - \frac{2}{{{L^2\sigma^2}}}\textbf{W}{(k)^\textrm{T}}{\bm{\Phi}} (k)\textbf{e}(k)\nonumber\\
 &=0,
\end{align}
where
\begin{align}
&\Gamma_1=\frac{1}{{{L^2\sigma^2}}}\sum\limits_{i = 1}^L {\sum\limits_{j = 1}^L {e{}_j(k){G_\sigma }} } (e{}_j(k) -  e{}_i(k)){\textbf{w}_j}{(k)^\textrm{T}}\\
&\Gamma_2= \frac{1}{{{L^2\sigma^2}}}\sum\limits_{i = 1}^L {\sum\limits_{j = 1}^L {e{}_i(k){G_\sigma }} } (e{}_j(k) - e{}_i(k)){\textbf{w}_j}{(k)^\textrm{T}}\\
& \Gamma_3= \frac{1}{{{L^2\sigma^2}}}\sum\limits_{i = 1}^L {\sum\limits_{j = 1}^L {e{}_j(k){G_\sigma }} } (e{}_j(k) - e{}_i(k)){\textbf{w}_i}{(k)^\textrm{T}}\\ &\Gamma_4=\frac{1}{{{L^2\sigma^2}}}\sum\limits_{i = 1}^L {\sum\limits_{j = 1}^L {e{}_i(k){G_\sigma }} } (e{}_j(k) - e{}_i(k)){\textbf{w}_i}{(k)^\textrm{T}}\\
 & [{\bm{\Phi}}(k)]_{ij}={G_\sigma }(e{}_j(k) - e{}_i(k))\\
& [{\bm{\Psi}}(k)]_{ij}={\sum\limits_{i = 1}^L {{G_\sigma }} (e{}_j(k) - e{}_i(k))}.
\end{align}
 From (\ref{error111}), $\textbf{x}{\rm{(}}k{\rm{)}}$ can be solved by a fixed-point iteration:
\begin{align}\label{fixedpointequation}
{\hat {\textbf{x}}}{\rm{(}}k{\rm{)}} &= g({\hat {\textbf{x}}}{\rm{(}}k{\rm{)}})\nonumber\\
&={\left( {\textbf{W}{{(k)}^\textrm{T}} {{\bm{\Lambda}}{\rm{(}}k{\rm{)}}\textbf{W}{\rm{(}}k{\rm{)}}} } \right)^{{\rm{ - }}1}}\left( {\textbf{W}{{(k)}^\textrm{T}} {{\bm{\Lambda}}{\rm{(}}k{\rm{)}}\textbf{d}{\rm{(}}k)} } \right),
\end{align}
where
\begin{align}
{\bm{\Lambda}}{\rm{(}}k{\rm{) = }}{\bm{\Psi}} {\rm{(}}k{\rm{)}} - {\bm{\Phi}} {\rm{(}}k{\rm{)}}=\left[ \begin{array}{l}
\begin{array}{*{20}{c}}
{{{\bm{\Lambda}}_x}(k)}&{{{\bm{\Lambda}}_{yx}}(k)}
\end{array}\\
\begin{array}{*{20}{c}}
{{{\bm{\Lambda}}_{xy}}(k)}&{{{\bm{\Lambda}}_y}(k)}
\end{array}
\end{array} \right],\label{35}
\end{align}
where ${\bm{\Lambda}}(k)\in \mathbb{R}^{L\times L}$.
The explicit expressions of
${{{\bm{\Lambda}}_x}(k)}\in \mathbb{R}^{n\times n}$, ${{{\bm{\Lambda}}_{xy}}(k)}\in \mathbb{R}^{m\times n}$, ${{{\bm{\Lambda}}_{yx}}(k)}\in \mathbb{R}^{n\times m}$ and ${{{\bm{\Lambda}}_y}(k)}\in \mathbb{R}^{m\times m}$
are
\begin{align}
&{{\bm{\Lambda}} _{{x}}}(k) = \left({{{\bm{\Lambda}} _{{{i,}}j}}(k)}\right)_{n\times n}=\left({\bm{\Phi}}_{i,j} (k)\right)_{n\times n}-\left({\bm{\Psi}}_{i,j} (k)\right)_{n\times n} \nonumber\\
&\,\,\,\,\,\,\,\,\,\,\,(i=1,2\cdots n;j=1,2\cdots n)\label{40} \\
&{{\bm{\Lambda}} _{{xy}}}(k) = \left( {{{\bm{\Lambda}} _{{{i,}}j}}(k)}\right)_{m\times n}=\left({\bm{\Phi}}_{i,j} (k)\right)_{m\times n}-\left({\bm{\Psi}}_{i,j} (k)\right)_{m\times n} \nonumber\\
 &\,\,\,\,\,\,\,\,\,\,\,(i=n+1,n+2\cdots n+m;j=1,2\cdots n)\label{41}\\
&{{\bm{\Lambda}} _{{yx}}}(k) = \left({{{\bm{\Lambda}} _{{{i,}}j}}(k)}\right)_{n\times m}=\left({\bm{\Phi}}_{i,j} (k)\right)_{n\times m}-\left({\bm{\Psi}}_{i,j} (k)\right)_{n\times m}\nonumber\\
&\,\,\,\,\,\,\,\,\,\,\,(i=1,2\cdots n;j=n+1,n+2\cdots n+m)\label{42}\\
&{{\bm{\Lambda}} _{{y}}}(k) =\left( {{{\bm{\Lambda}} _{{{i,}}j}}(k)} \right)_{m\times m}=\left({\bm{\Phi}}_{i,j} (k)\right)_{m\times m}-\left({\bm{\Psi}}_{i,j} (k)\right)_{m\times m}\nonumber\\ &\,\,\,\,\,\,\,\,\,\,\,(i=n+1,n+2\cdots n+m; \nonumber\\
&\,\,\,\,\,\,\,\,\,\,\,\,j=n+1,n+2\cdots n+m).\label{43}
\end{align}

According to  Eqs. (\ref{21}), (\ref{22}) and (\ref{35}), we arrive at
\begin{align}\label{simplification}
\begin{array}{l}
\left( {\textbf{W}{{(k)}^\textrm{T}} {{\bm{\Lambda}}{\rm{(}}k{\rm{)}}\textbf{W}{\rm{(}}k{\rm{)}}} } \right)\\
 = \left[{\bm{\Theta}}_p^{ - 1}{(k|k - 1)^\textrm{T}}{{\bm{\Lambda}}_x} + {\textbf{H}^\textrm{T}}{\bm{\Theta}}_r^{ - 1}{(k)^\textrm{T}}{{\bm{\Lambda}}_{xy}}\right]{\bm{\Theta}}_p^{ - 1}(k|k - 1)\\
+\left[{\bm{\Theta}}_p^{ - 1}{(k|k - 1)^\textrm{T}}{{\bm{\Lambda}}_{yx}} + {\textbf{H}^\textrm{T}}{\bm{\Theta}}_r^{ - 1}{(k)^\textrm{T}}{{\bm{\Lambda}}_y}\right]{\bm{\Theta}}_r^{ - 1}(k)\textbf{H}\\
\left( {\textbf{W}{{(k)}^\textrm{T}} {{\bm{\Lambda}}{\rm{(}}k{\rm{)}}\textbf{D}{\rm{(}}k{\rm{)}}} } \right)\\
 = \left[{\bm{\Theta}}_p^{ - 1}{(k|k - 1)^\textrm{T}}{{\bm{\Lambda}}_x} + {\textbf{H}^\textrm{T}}{\bm{\Theta}}_r^{ - 1}{(k)^\textrm{T}}{{\bm{\Lambda}}_{xy}}\right]{\bm{\Theta}}_p^{ - 1}(k|k - 1)\\
{\hat {\textbf{x}}}{\rm{(}}k|k - 1{\rm{) + }}\left[{\bm{\Theta}}_p^{ - 1}{(k|k - 1)^\textrm{T}}{{\bm{\Lambda}}_{yx}} + {\textbf{H}^\textrm{T}}{\bm{\Theta}}_r^{ - 1}{(k)^\textrm{T}}{{\bm{\Lambda}}_y}\right]\\
{\bm{\Theta}}_r^{ - 1}(k)\textbf{y}{\rm{(}}k{\rm{)}}.
\end{array}
\end{align}
By (\ref{simplification}), the Eq. (\ref{fixedpointequation}) can be rewritten as
\begin{align}\label{textbfx}
{\hat {\textbf{x}}}{\rm{(}}k{\rm{)}}
 &={\left( {{\bm{\Omega}_1} + {\bm{\Omega}_2\bm{\Omega}_3}} \right)^{{\rm{ - }}1}}\left( {{\bm{\Omega}_1}{\hat {\textbf{x}}}{\rm{(}}k|k - 1{\rm{) + }}{\bm{\Omega}_2}\textbf{y}{{(k)}}} \right)\nonumber\\
 &= {\left( {{\bm{\Omega}_1} + {\bm{\Omega}_2}\bm{\Omega}_3} \right)^{{\rm{ - }}1}}{\bm{\Omega}_1}{\hat {\textbf{x}}}{\rm{(}}k|k - 1)\nonumber\\
  &+ {\left( {{\bm{\Omega}_1} + {\bm{\Omega}_2}\bm{\Omega}_3} \right)^{{\rm{ - }}1}}{\bm{\Omega}_2}\textbf{y}(k),
\end{align}
where
\begin{align}
\begin{array}{l}
{\bm{\Omega}_1}{\rm{ = }}\left[{\bm{\Theta}}_p^{ - 1}{(k|k - 1)^\textrm{T}}{{\bm{\Lambda}}_x} + {\textbf{H}^\textrm{T}}{\bm{\Theta}}_r^{ - 1}{(k)^\textrm{T}}{{\bm{\Lambda}}_{xy}}\right]{\bm{\Theta}}_p^{ - 1}(k|k - 1)\\
{\bm{\Omega}_2}{\rm{ = }}\left[{\bm{\Theta}}_p^{ - 1}{(k|k - 1)^\textrm{T}}{{\bm{\Lambda}}_{yx}} + {\textbf{H}^\textrm{T}}{\bm{\Theta}}_r^{ - 1}{(k)^\textrm{T}}{{\bm{\Lambda}}_y}\right]{\bm{\Theta}}_r^{ - 1}(k)\nonumber\\
{\bm{\Omega}_3}{\rm{ = }}\textbf{H}.
\end{array}
\end{align}
By using the matrix inversion lemma
\begin{align}
\begin{array}{l}
{\left( {\textbf{A} + \textbf{BCD}} \right)^{{\rm{ - }}1}}\\
{\rm{ = }}{\textbf{A}^{{\rm{ - }}1}} - {\textbf{A}^{{\rm{ - }}1}}{\textbf{B}}{{\rm{(\textbf{C}}^{-1}} + \textbf{D}{\textbf{A}^{{\rm{ - }}1}}{\textbf{B}}{\rm{)}}^{ - 1}}\textbf{D}{\textbf{A}^{{\rm{ - }}1}},
\end{array}
\end{align}
with the identifications
\begin{align}
{\bm{\Omega}_1}\rightarrow \textbf{A},{\bm{\Omega}_2}\rightarrow \textbf{B},
\textbf{I}_L \rightarrow \textbf{C},{\bm{\Omega}_3}\rightarrow \textbf{D},
\end{align}
one can reformulate (\ref{textbfx}) as
\begin{align}
\begin{array}{l}
{\hat {\textbf{x}}}{\rm{(}}k{\rm{)}}
{\rm{ = }}{\hat {\textbf{x}}}{\rm{(}}k|k - 1{\rm{) + }}\overline {\textbf{K}}{\rm{(}}k{\rm{)}}\left( {\textbf{y}{\rm{(}}k{\rm{)}} - \textbf{H}{\hat {\textbf{x}}}{\rm{(}}k|k - 1{\rm{)}}} \right),
\end{array}
\end{align}
where
\begin{align}
\begin{array}{l}
\overline {\textbf{K}}(k) =
\left[ \begin{array}{l}
\overline {\textbf{P}}(k|k - 1) + {\textbf{H}^\textrm{T}}{{\overline {\textbf{P}}}_{xy}}(k|k - 1) \\
+\left( {\overline {\textbf{P}}}_{yx}(k|k - 1) + {\textbf{H}^\textrm{T}}\overline {\textbf{R}}(k) \right)\textbf{H}
\end{array} \right]^{-1}\\
\,\,\,\,\,\,\,\,\,\,\,\,\,\,\times \left( {{{\overline {\textbf{P}}}_{yx}}(k|k - 1){\rm{ + }}{\textbf{H}^\textrm{T}}\overline {\textbf{R}}(k)} \right)\\
\overline {\textbf{P}}(k|k - 1) = {\bm{\Theta}}_p^{ - 1}{(k|k - 1)^\textrm{T}}{{\bm{\Lambda}}_x}(k){\bm{\Theta}}_p^{ - 1}(k|k - 1)\\
{{\overline {\textbf{P}}}_{xy}}(k|k - 1) = {\bm{\Theta}}_r^{ - 1}{(k)^\textrm{T}}{{\bm{\Lambda}}_{xy}}(k){\bm{\Theta}}_p^{ - 1}(k|k - 1)\\
{{\overline {\textbf{P}}}_{yx}}(k|k - 1) = {\bm{\Theta}}_p^{ - 1}{(k|k - 1)^\textrm{T}}{{\bm{\Lambda}}_{yx}}(k){\bm{\Theta}}_r^{ - 1}(k)\\
\overline {\textbf{R}}(k) = {\bm{\Theta}}_r^{ - 1}{(k)^\textrm{T}}{{\bm{\Lambda}}_y}(k){\bm{\Theta}}_r^{ - 1}(k).
\end{array}
\end{align}

Then, the posterior covariance matrix can be updated by
\begin{align}\label{pca}
\textbf{P}(k) = &\left[\textbf{I} - \overline {\textbf{K}}(k)\textbf{H}\right]\textbf{P}(k{\rm{|}}k - {\rm{1}}){\left[\textbf{I} - \overline {\textbf{K}}(k)\textbf{H}\right]^\textrm{T}} \nonumber\\
&+ \overline {\textbf{K}}(k)\textbf{R}(k)\overline {\textbf{K}}{(k)^\textrm{T}}.
\end{align}
With the above derivations, the proposed MEE-KF algorithm can be summarized as \textbf{Algorithm 1}.
\small
\begin{algorithm}
\caption{Minimum Error Entropy Kalman Filter (MEE-KF)} \label{MEEKF}
\textbf{Step 1:} Initialize the state priori estimate ${\hat {\textbf{x}}} (1|0)$ and state prediction error covariance matrix $\textbf{P}(1|0)$; set a proper kernel size $\sigma$ and a small positive number $\varepsilon$.\\
 \textbf{Step 2:} Use Eqs. (\ref{Predict1}) and (\ref{Predict2}) to obtain ${\hat {\textbf{x}}}(k|k- 1)$ and $\textbf{P}(k|k-1)$, respectively;  use the Cholesky decomposition of ${\bm{\Theta}}(k)$ to obtain ${{\bm{\Theta}}_p}{(k|k-1)}$ and ${{\bm{\Theta}}_r}(k)$; use Eqs. (\ref{21}) and (\ref{22}) to obtain ${ {\textbf{d}}}(k)$ and $\textbf{W}(k-1)$, respectively.\\
 \textbf{Step 3:} Let $t = 1$ and ${\hat {\textbf{x}}}(k)_0={\hat {\textbf{x}}}(k|k - 1)$, where ${\hat {\textbf{x}}}(k)_t$ denotes the estimated state at the fixed-point iteration $t$.\\
\textbf{Step 4:}
Use available measurements $\{\textbf{y}(k)\}^{N}_{k=1}$ to
update:\\
\begin{align}
&{\hat {\textbf{x}}}(k)_t = {\hat {\textbf{x}}}(k|k - 1) + \widetilde{{\textbf{K}}}(k)[\textbf{y}(k) - \textbf{H}{\hat {\textbf{x}}}(k|k - 1)]\label{5011}
\end{align}
with
\begin{align}
&\widetilde{\textbf{e}}_i(k)=\textbf{d}_i(k)-\textbf{w}(k)\hat{\textbf{x}}(k)_{t-1} \label{49}\\
&\widetilde{\bm{\Lambda}}{\rm{(}}k{\rm{) = }}\widetilde{\bm{\Psi}} {\rm{(}}k{\rm{)}} - \widetilde{\bm{\Phi}} {\rm{(}}k{\rm{)}}=\left[ \begin{array}{l}
\begin{array}{*{20}{c}}
{{\widetilde{\bm{\Lambda}}_x}(k)}&{{\widetilde{\bm{\Lambda}}_{yx}}(k)}
\end{array}\\
\begin{array}{*{20}{c}}
{{\widetilde{\bm{\Lambda}}_{xy}}(k)}&{{\widetilde{\bm{\Lambda}}_y}(k)}
\end{array}
\end{array} \right]\label{new44}\\
&\widetilde {\textbf{P}}(k|k - 1) = {\bm{\Theta}}_p^{ - 1}{(k|k - 1)^\textrm{T}}{\widetilde{\bm{\Lambda}}_x}(k){\bm{\Theta}}_p^{ - 1}(k|k - 1)\label{44}\\
&{{\widetilde{\textbf{P}}}_{xy}}(k|k - 1) = {\bm{\Theta}}_r^{ - 1}{(k)^\textrm{T}}{\widetilde{\bm{\Lambda}}_{xy}}(k){\bm{\Theta}}_p^{ - 1}(k|k - 1)\label{45}\\
&{{\widetilde {\textbf{P}}}_{yx}}(k|k - 1) = {\bm{\Theta}}_p^{ - 1}{(k|k - 1)^\textrm{T}}{\widetilde{\bm{\Lambda}}_{yx}}(k){\bm{\Theta}}_r^{ - 1}(k)\label{46}\\
&\widetilde {\textbf{R}}(k) = {\bm{\Theta}}_r^{ - 1}{(k)^\textrm{T}}{\widetilde{\bm{\Lambda}}_y}(k){\bm{\Theta}}_r^{ - 1}(k)\label{47}\\
&\widetilde {\textbf{K}}(k) =
\left[ \begin{array}{l}
\widetilde {\textbf{P}}(k|k - 1) + {\textbf{H}^\textrm{T}}{{\widetilde {\textbf{P}}}_{xy}}(k|k - 1) \\
+\left( {\widetilde {\textbf{P}}}_{yx}(k|k - 1) + {\textbf{H}^\textrm{T}}\widetilde {\textbf{R}}(k) \right)\textbf{H}
\end{array} \right]^{-1}\nonumber\\
&\,\,\,\,\,\,\,\,\,\,\,\,\,\,\times \left( {{{\widetilde {\textbf{P}}}_{yx}}(k|k - 1){\rm{ + }}{\textbf{H}^\textrm{T}}\widetilde {\textbf{R}}(k)} \right).\label{48}
\end{align}
\textbf{Step 5:}
Compare $\hat {\textbf{x}}{{(k)}_t}$ and $\hat {\textbf{x}}{{(k)}_{t - 1}}$
\begin{align}
\frac{{||\hat {\textbf{x}}{{(k)}_t} - \hat {\textbf{x}}{{(k)}_{t - 1}}||}}{{||\hat {\textbf{x}}(k)_{t - 1}||}} \le \varepsilon.
\end{align}
If the above condition holds, set $\hat {\textbf{x}}(k)=\hat {\textbf{x}}(k)_t$ and continue to \textbf{Step 6}. Otherwise, $t + 1\rightarrow t$, and return to \textbf{Step 4}.\\
\textbf{Step 6:}
Update $k + 1\rightarrow k$ and the posterior error covariance matrix by
\begin{align}
&\textbf{P}(k){\rm{ = [}}\textbf{I} - \widetilde{{\textbf{K}}}(k)\textbf{H}]\textbf{P}(k|k - 1){{\rm{[}}\textbf{I} - \widetilde{{\textbf{K}}}(k)\textbf{H}]^\textrm{T}}\nonumber\\
&\,\,\,\,\,\,\,\,\,\,\,\,\,\,\,+ \widetilde{{\textbf{K}}}(k)\textbf{R}(k)\widetilde{{\textbf{K}}}^{\textrm{T}}(k),\label{5911}
\end{align}
and return to \textbf{Step 2}.
\end{algorithm}
\normalsize

%

\section{Computational Complexity}
This section provides the comparison of the computational complexities of KF, maximum correntropy Kalman filter (MCKF) \cite{KFMCchen} and MEE-KF in terms of the floating point operations.

The KF updates with Eqs. (\ref{Predict1})-(\ref{update2}), and the corresponding floating point operations are given in Table \ref{firsttable1}. From Table \ref{firsttable1},  we can conclude that the computational complexity of KF is
\begin{align}
C_{KF}=8n^3+10n^2m-n^2+6nm^2-n+O(m^3).
\end{align}

According to \cite{KFMCchen}, the computational complexity of MCKF is
\begin{align}
C_{MCKF}&=(2T + 8)n^3 + (4T + 6)n^2m + (2T-1)n^2\nonumber\\
&+(4T + 2)nm^2 + (3T-1)nm + (4T-1)n\nonumber\\
&+2Tm^3 + 2Tm + TO(n^3) + 2TO(m^3),
\end{align}
where $T$ denotes the fixed-point iteration number, which is relatively small in general as shown in simulations in Section VII.

The updates of MEE-KF involve Eqs. (\ref{Predict1}), (\ref{Predict2}), (\ref{5011})-(\ref{48}) and (\ref{5911}), and the corresponding floating point operations are shown in Table \ref{firsttable1}. According to Table \ref{firsttable1}, the computational complexity of MEE-KF is
\begin{align}
C_{MEE-KF} &= (7T+8)n^3+(7T)m^3+(19T+6)n^2m-n^2\nonumber\\
&+(15T+2)nm^2+TO(n^3)+TO(mn) \nonumber\\
&+Tm+(5T-1)n+(7T-1)nm\nonumber\\
&+TO(m^2)+2TO(m^3).
\end{align}

The MEE-KF has an additional computational burden induced by the error entropy functions in comparison to KF, and has a slightly higher computational complexity than MCKF. In the sense of order of magnitude, the computational complexities of the MEE-KF, MCKF and KF have no significant difference.
\begin{table}[!t]
\caption{Computational Complexities of Some Equations}
\begin{center}
\begin{tabular}{l p{3cm}  p{2.8cm}}
\hline
\multirow{1}{*}{Equation}&{ Addition/subtraction and multiplication} &{Division, matrix inversion, Cholesky decomposition and exponentiation }\\
\cline{1-3}
\multirow{1}{*}{(\ref{Predict1})}&{$2n^2-n$}&$0$\\
\multirow{1}{*}{(\ref{Predict2})}&$4n^3-n^2$&$0$\\
\multirow{1}{*}{(\ref{update1})}&$4nm$&$0$\\
\multirow{1}{*}{(\ref{update3})}&$4n^2m+4nm^2- 3nm$&${O({m^{3}})}$\\
\multirow{1}{*}{(\ref{update2})}&$4n^3+6n^2m-2n^2+2nm^2-nm$&$0$\\
\multirow{1}{*}{(\ref{5011})}&$4nm$&${O({m^{2}})}+{O(mn)}$\\
\multirow{1}{*}{(\ref{49})}&$2n$&$0$\\
\multirow{1}{*}{(\ref{new44})}&$3(m+n)^3+(m+n)^2$&$(m+n)^2$\\
\multirow{1}{*}{(\ref{44})}&$4n^3-2n^2$&$n+{O({n^{3}})}$\\
\multirow{1}{*}{(\ref{45})}&$2n^2m+2nm^2-2nm$&$0$\\
\multirow{1}{*}{(\ref{46})}&$2n^2m+2nm^2-2nm$&$0$\\
\multirow{1}{*}{(\ref{47})}&$4m^3-2m^2$&$m+{O({m^{3}})}$\\
\multirow{1}{*}{(\ref{48})}&$6n^2m+2nm^2-nm$&${O({m^{3}})}$\\
\multirow{1}{*}{(\ref{5911})}&$4n^3+6n^2m-2n^2+2nm^2-nm$&$0$\\
\cline{1-3}
\end{tabular}
\label{firsttable1}
\end{center}
\end{table}
\section{Convergence Issue}
This section provides a sufficient condition to ensure the convergence of the fixed point iterations in MEE-KF, where the proof is similar to \cite{CMEE} and thus will not be provided here.

First, from Eq. (\ref{fixedpointequation}), we can rewrite
\begin{align}
g(\textbf{x}(k)) = \textbf{M}_{\textbf{w}\textbf{w}}^{ - 1}{\textbf{M}_{\textbf{dw}}}
\end{align}
with $
\textbf{M}_{\textbf{ww}}^{ - 1} = \sum\limits_{i = 1}^L {\sum\limits_{j = 1}^L \begin{array}{l}
{G_\sigma }(e{}_j(k) - e{}_i(k))[{\textbf{w}_j}(k) - {\textbf{w}_i}(k)]\\
 \times {[{\textbf{w}_j}(k) - {\textbf{w}_i}(k)]^{\rm{T}}}
\end{array} }$ and $
{\textbf{M}_{\textbf{dw}}} = \sum\limits_{i = 1}^L {\sum\limits_{j = 1}^L \begin{array}{l}
{G_\sigma }(e{}_j(k) - e{}_i(k))][{{d}_j}(k) - {{d}_i}(k)]\\
 \times [{\textbf{w}_j}(k) - {\textbf{w}_i}(k)]
\end{array}}$.

Thus, a $n\times n$ Jacobian matrix of $g(\textbf{x}(k))$ with respect to $\textbf{x}(k)$ gives
\begin{align}
\begin{array}{l}
{\nabla _{\textbf{x}(k)}}g(\textbf{x}(k)) = \dfrac{\partial }{{\partial {\textbf{x}(k)}}}\textbf{M}_{\textbf{ww}}^{ - 1}{\textbf{M}_{\textbf{dw}}}\\
 = \left[ {\frac{{\partial g(\textbf{x}(k))}}{{\partial {{\textbf{x}}_1(k)}}}{\kern 1pt} {\kern 1pt} {\kern 1pt} {\kern 1pt} {\kern 1pt} {\kern 1pt} {\kern 1pt} {\kern 1pt} \frac{{\partial g(\textbf{x}(k))}}{{\partial {{\textbf{x}}_2(k)}}} \cdots \frac{{\partial g(\textbf{x}(k))}}{{\partial {{\textbf{x}}_L}}}} \right]\\
 =  - \dfrac{{\textbf{M}_{\textbf{ww}}^{ - 1}}}{{2{L^2}{\sigma ^2}}}\left( {\sum\limits_{i = 1}^L {\sum\limits_{j = 1}^L \begin{array}{l}
{\xi _1}(k)[{\textbf{w}_j}(k) - {\textbf{w}_i}(k)]\\
 \times {[{\textbf{w}_j}(k) - {\textbf{w}_i}(k)]^\textrm{T}}
\end{array} } } \right)g(\textbf{x}(k))\\
 + \dfrac{{\textbf{M}_{\textbf{ww}}^{ - 1}}}{{2{L^2}{\sigma ^2}}}\left( {\sum\limits_{i = 1}^L {\sum\limits_{j = 1}^L \begin{array}{l}
{\xi _1}(k)[{d_j}(k) - {d_i}(k)]\\
 \times [{\textbf{w}_j}(k) - {\textbf{w}_i}(k)]
\end{array} } } \right),
\end{array}
\end{align}
with ${\xi _1}(k) = ({e_j}(k) - {e_i}(k))({\textbf{w}_j}(k) - {\textbf{w}_i}(k)){G_\sigma }(e(j) - e(i))$.
Define $|| \cdot ||_p$ as an $l_p$-norm ($p\geq 1$) of a vector or an induced norm of a matrix as $||\textbf{A}|{|_p} = \mathop {\max }\limits_{{\rm{||\textbf{W}}}|{|_p} \ne 0} ||\textbf{AW}|{|_p}/||\textbf{W}|{|_p}$. According to the proof in \cite{CMEE}, the following theorem holds.

\begin{thm}
If the kernel size satisfies $\sigma \geq \textrm{max}\{\sigma_1,\sigma_2\}$, we have
\begin{align}
\left\{ \begin{array}{l}
||g(\textbf{x}(k))|{|_1} \le \beta \\
||{\nabla _{\textbf{x}(k)}}g(\textbf{x}(k))|{|_1}  \le \alpha < 1,
\end{array} \right.
\end{align}
where $\beta  > \rho  = \frac{{\sqrt n \sum\limits_{i = 1}^L {\sum\limits_{j = 1}^L \begin{array}{l}
[{d_j}(k) - {d_i}(k)]\\
 \times [{\textbf{w}_j}(k) - {\textbf{w}_i}(k)]
\end{array} } }}{{{\lambda _{\min }}\left[ {\sum\limits_{i = 1}^L {\sum\limits_{j = 1}^L \begin{array}{l}
[{\textbf{w}_j}(k) - {\textbf{w}_i}(k)]\\
 \times {[{\textbf{w}_j}(k) - {\textbf{w}_i}(k)]^\textrm{T}}
\end{array} } } \right]}}$; ${\lambda _{\min }}$ is the minimum eigenvalue of the matrix $\textbf{M}_{\textbf{ww}}$; $ \sigma_1 $ and $\sigma_2 $  are the solutions of $||g(\textbf{x}(k))|{|_1}= \beta$ and $||{\nabla _{\textbf{x}(k)}}g(\textbf{x}(k))|{|_1}= \alpha$, respectively, i.e.,
\begin{align}
||g(\textbf{x}(k))|{|_1} = \frac{{\sqrt n \sum\limits_{i = 1}^L {\sum\limits_{j = 1}^L \begin{array}{l}
[{d_j}(k) - {d_i}(k)]\\
 \times [{\textbf{w}_j}(k) - {\textbf{w}_i}(k)]
\end{array} } }}{{{\lambda _{\min }}\left[ {\sum\limits_{i = 1}^L {\sum\limits_{j = 1}^L \begin{array}{l}
{\xi _2}(k)[{\textbf{w}_j}(k) - {\textbf{w}_i}(k)]\\
 \times {[{\textbf{w}_j}(k) - {\textbf{w}_i}(k)]^\textrm{T}}
\end{array} } } \right]}}
\end{align}
with ${\xi _2}(k) = {G_\sigma }(\beta ||{\textbf{w}_j}(k) - {\textbf{w}_i}(k)|{|_1} + |{d_j}(k) - {d_i}(k)|)$, and
\begin{align}
||{\nabla _{\textbf{x}(k)}}g(\textbf{x}(k))|{|_1} = \frac{{\gamma \sqrt {{n}} }}{{{\sigma ^2}{\lambda _{\min }}\left[ {\sum\limits_{i = 1}^L {\sum\limits_{j = 1}^L \begin{array}{l}
{\xi _2}(k)[{\textbf{w}_j}(k) - {\textbf{w}_i}(k)]\\
 \times {[{\textbf{w}_j}(k) - {\textbf{w}_i}(k)]^\textrm{T}}
\end{array} } } \right]}}
\end{align}
with $\gamma =\sum\limits_{i = 1}^L {\sum\limits_{j = 1}^L \begin{array}{l}
(\beta ||{\textbf{w}_j}(k) - {\textbf{w}_i}(k)|{|_1} + |{d_j}(k) - {d_i}(k)|)\\
||{\textbf{w}_j}(k) - {\textbf{w}_i}(k)|{|_1}(\beta ||[{\textbf{w}_j}(k) - {\textbf{w}_i}(k)]\\
{[{\textbf{w}_j}(k) - {\textbf{w}_i}(k)]^\textrm{T}}|{|_1} + [{d_j}(k) - {d_i}(k)]\\
||{\textbf{w}_j}(k) - {\textbf{w}_i}(k)|{|_1})
\end{array} }$.
\end{thm}
According to \textbf{Theorem 1} and Banach Fixed Point Theorem, the fixed-point algorithm induced by (\ref{fixedpointequation})  can be guaranteed to converge to a unique fixed point in the range $\textbf{x}(k)\in \{||\textbf{x}(k)||_1\leq \beta \}$ provided that the kernel size $\sigma$ is larger than a certain value and an initial state vector satisfies $||\textbf{x}(k)_0||_1\leq \beta$.
\section{Minimum Error Entropy Extended Kalman Filter}
For a nonlinear system, the state and measurement equations can be described by
\begin{align}
&\textbf{x}(k) = \textbf{f}(\textbf{x}(k - 1))+\textbf{q}(k - 1)\label{stateEKF}\\
&\textbf{y}(k)=\textbf{h}(\textbf{x}(k)) + \textbf{r}(k),\label{measurementEKF}
\end{align}
where $\textbf{f}(\cdot)$ and $\textbf{h}(\cdot)$ are the state-transition function and measurement (observation) function, respectively. The EKF is an important tool for dealing with the state estimation of a nonlinear system via approximating the nonlinear system by linear model. In EKF, the nonlinear functions ${\textbf{f}}({\textbf{x}}(k - 1))$ and $\textbf{h}(\textbf{x}(k))$ can be approximated by the first-order Taylor series expansion at ${\hat {\textbf{x}}}(k)$ and ${\hat {\textbf{x}}}(k|k - 1)$,  i.e.,
\begin{align}
&\textbf{f}(\textbf{x}(k-1))\approx \textbf{f}({\hat {\textbf{x}}}(k - 1))+\textbf{F}(k-1)(\textbf{x}(k-1)-{\hat {\textbf{x}}}(k - 1))\label{firstorderf}\\
&\textbf{h}(\textbf{x}(k))\approx \textbf{h}({\hat {\textbf{x}}}(k|k - 1))+\textbf{H}(k)(\textbf{x}(k)-{\hat {\textbf{x}}}(k|k - 1))\label{firstorder},
\end{align}
where $\textbf{F}(k-1)$ and $\textbf{H}(k)$ are the Jacobian matrices of $\textbf{f}(\cdot)$ and $\textbf{h}(\cdot)$, i.e.,
\begin{align}
&\textbf{F}(k-1) = {\left. {\frac{{\partial {\textbf{f}(\textbf{x}(k-1))}}}{{\partial \textbf{x}}}} \right|_{\textbf{x} = \hat {\textbf{x}}(k)}}\label{Jacobian1}\\
&\textbf{H}(k) = {\left. {\frac{{\partial {\textbf{h}({ {\textbf{x}}}(k))}}}{{\partial \textbf{x}}}} \right|_{\textbf{x} = \hat {\textbf{x}}(k|k - 1)}}\label{Jacobian2}.
\end{align}

By substituting  (\ref{firstorderf}) and (\ref{firstorder}) into  (\ref{stateEKF}) and (\ref{measurementEKF}),  the prediction and update equations of EKF can be obtained:
\begin{align}
&{\hat {\textbf{x}}}(k|k - 1){\rm{ = }}\textbf{f}({\hat {\textbf{x}}}(k - 1))\label{Predict1EKF}\\
&\textbf{P}(k|k - 1){\rm{ = }}\textbf{F}(k-1)\textbf{P}(k - 1){\textbf{F}^{\rm{T}}}(k-1){\rm{ + }}\textbf{Q}(k - 1)\label{Predict2EKF}\\
&\textbf{K}(k){\rm{ = }}\textbf{P}({{k|}}k - 1){\textbf{H}^\textrm{T}(k)}[\textbf{H}(k)\textbf{P}({{k|}}k - 1){\textbf{H}^\textrm{T}(k)} + \textbf{R}(k)]^{-1}\label{update3EKF}\\
&{\hat {\textbf{x}}}(k) = {\hat {\textbf{x}}}(k|k - 1) + {\textbf{K}}(k)[\textbf{y}(k) - \textbf{h}({\hat {\textbf{x}}}(k|k - 1))]\label{update1EKF}\\
&\textbf{P}(k){\rm{ = [}}\textbf{I} - {\textbf{K}}(k)\textbf{H}(k)]\textbf{P}(k|k - 1){{\rm{[}}\textbf{I} - {\textbf{K}}(k)\textbf{H}(k)]^\textrm{T}}\nonumber\\
&\,\,\,\,\,\,\,\,\,\,\,\,\,\,\,+ {\textbf{K}}(k)\textbf{R}(k){\textbf{K}^{\textrm{T}}}(k),\label{update2EKF}
\end{align}
where ${\textbf{K}}(k)$ is the gain of EKF. Similar to MEE-KF and maximum correntropy extended Kalman filter (MCEKF) \cite{MCEKF}, the augmented model of MEE-EKF should be established. By substituting  (\ref{firstorder}) into  (\ref{measurementEKF}), the measurement equation is approximated by
\begin{align}\label{78}
\textbf{y}(k)&\approx \textbf{h}({\hat {\textbf{x}}}(k|k - 1))\nonumber\\
&+\textbf{H}(k)(\textbf{x}(k)-{\hat {\textbf{x}}}(k|k - 1)) + \textbf{r}(k).
\end{align}
Then, replacing $\textbf{y}(k)$ in (\ref{augmentedmodel}) with that in (\ref{78}), one can obtain the following augmented model:
\begin{align}\label{augmentedmodel1}
\begin{array}{c}
\left[ \begin{array}{l}
{\hat {\textbf{x}}}(k|k - 1)\\
\textbf{y}(k)-\textbf{h}({\hat {\textbf{x}}}(k|k - 1))+\textbf{H}(k){\hat {\textbf{x}}}(k|k - 1)
\end{array} \right] \\
= \left[ \begin{array}{l}
\textbf{I}_{n}\\
\textbf{H}(k)
\end{array} \right]\textbf{x}(k) +\mu(k),
\end{array}
\end{align}
where $\mu(k)$ has the same form with (\ref{20}).
Similar to (\ref{equation23}), we have
\begin{align}\label{80}
\textbf{d}(k)= \textbf{W}(k)\textbf{x}(k) + \textbf{e}(k),
\end{align}
where
\begin{align}
&\textbf{d}(k){\rm{ = }}{{\bm{\Theta}}^{{\rm{ - }}1}}(k)\left[ \begin{array}{l}
{\hat {\textbf{x}}}(k|k - 1)\\
\textbf{y}(k)-\textbf{h}({\hat {\textbf{x}}}(k|k - 1))+\textbf{H}(k){\hat {\textbf{x}}}(k|k - 1)
\end{array} \right]\label{2285}\\
&\textbf{W}(k) = {{\bm{\Theta}}^{{\rm{ - }}1}}(k)\left[ \begin{array}{l}
\textbf{I}_{n}\\
\textbf{H}(k)
\end{array} \right]\label{2286}.
\end{align}
The forms of ${\bm{\Theta}}(k)$ and $\textbf{e}(k)$ are the same as Eqs. (\ref{211}) and (\ref{27}), respectively.
Therefore, using the Eq. (\ref{80}) and the derivation similar to that of MEE-KF, one can obtain the MEE-EKF, as provided in \textbf{Algorithm 2}.
\small
\begin{algorithm}[!t]
\caption{Minimum Error Entropy Extended Kalman Filter (MEE-EKF)} \label{MEEKF}
\textbf{Step 1:} Similar to the  \textbf{Step 1} in \textbf{Algorithm 1}.\\
 \textbf{Step 2:} Use Eqs. (\ref{Jacobian1}), (\ref{Predict1EKF}), and (\ref{Predict2EKF}) to obtain ${\hat{\textbf{x}}}(k|k- 1)$ and $\textbf{P}(k|k-1)$; use the Cholesky decomposition of ${\bm{\Theta}}(k)$ to obtain ${{\bm{\Theta}}_p}{(k|k-1)}$ and ${{\bm{\Theta}}_r}(k)$; use Eqs. (\ref{2285}) and (\ref{2286}) to obtain ${ {\textbf{d}}}(k)$ and $\textbf{W}(k-1)$, respectively.\\
 \textbf{Step 3:} Similar to the  \textbf{Step 3} in \textbf{Algorithm 1}.\\
 \textbf{Step 4:}
Use the available measurements $\{\textbf{y}(k)\}^{N}_{k=1}$ to
update:\\
\begin{align}
&{\hat {\textbf{x}}}(k)_t = {\hat {\textbf{x}}}(k|k - 1) + \widetilde{{\textbf{K}}}(k)[\textbf{y}(k) - \textbf{h}({\hat {\textbf{x}}}(k|k - 1))]\label{3911}\\
&\widetilde{\textbf{e}}_i(k)=\textbf{d}_i(k)-\textbf{w}(k)\hat{\textbf{x}}(k)_{t-1}\label{4911}\\
&\widetilde {\textbf{K}}(k) =
\left[ \begin{array}{l}
\widetilde{\textbf{P}}(k|k - 1) + {\textbf{H}^\textrm{T}(k)}{{\widetilde {\textbf{P}}}_{xy}}(k|k - 1) \\
+\left( {\widetilde {\textbf{P}}}_{yx}(k|k - 1) + {\textbf{H}^\textrm{T}(k)}\widetilde {\textbf{R}}(k) \right)\textbf{H}(k)
\end{array} \right]^{-1}\nonumber\\
&\,\,\,\,\,\,\,\,\,\,\,\,\,\,\times \left( {{{\widetilde {\textbf{P}}}_{yx}}(k|k - 1){\rm{ + }}{\textbf{H}^\textrm{T}(k)}\widetilde {\textbf{R}}(k)} \right)\label{4811},
\end{align}
where the forms of
$\widetilde{\bm{\Lambda}}(k)$, $\widetilde {\textbf{P}}(k|k - 1)$, ${{\widetilde {\textbf{P}}}_{xy}}(k|k - 1)$, ${{\widetilde {\textbf{P}}}_{yx}}(k|k - 1)$ and $\widetilde {\textbf{R}}(k)$ are the same as Eqs. (\ref{new44})-(\ref{47}), respectively.\\
\textbf{Step 5:}
Similar to the \textbf{Step 1} in \textbf{Algorithm 1}.\\
 \textbf{Step 6:}
Update $k + 1\rightarrow k$ and the posterior covariance matrix:
\begin{align}
&\textbf{P}(k){\rm{ = [}}\textbf{I} - \widetilde{{\textbf{K}}}(k)\textbf{H}(k)]\textbf{P}(k|k - 1){{\rm{[}}\textbf{I} - \widetilde{{\textbf{K}}}(k)\textbf{H}(k)]^\textrm{T}}\nonumber\\
&\,\,\,\,\,\,\,\,\,\,\,\,\,\,\,+ \widetilde{{\textbf{K}}}(k)\textbf{R}(k)\widetilde{{\textbf{K}}}^{\textrm{T}}(k)\label{50}
\end{align}
and return to \textbf{Step 2}.
\end{algorithm}
\normalsize
\begin{table*}
\caption{Estimation Results of Different Algorithms Under Different Noises}
\begin{center}
\begin{tabular}{l l l l l l }
\hline
{Noises}&
{Algorithms}&{MSE of x1 }&{MSE of x2 }&{MSE of x3 }&{MSE of x4 }\\
\cline{1-6}
{{Gaussian noise}}&{KF}&{${0.0762}\pm{0.0087}$}&{${0.0762}\pm{0.0026}$}&{$ {0.0626}\pm0.0089$}&{$0.0593\pm0.0027$}\\
{}&{MCKF}&{$0.0762\pm0.0074$}&{$0.0762\pm0.0027$}&{$ 0.0627\pm0.0080$}&{$0.0593\pm0.0059$}\\
{}&{MEE-KF}&{${0.0791}\pm{0.0042}$}&{$0.0789\pm0.0045$}&{${0.0711}\pm{0.0010}$}&{${0.0725}\pm{0.0085}$}\\
{{Gaussian noise}}&{KF}&{$ 0.5011\pm0.0502$}&{$0.4868\pm0.0783$}&{$0.1595\pm0.0812$}&{$0.1396\pm0.0189$}\\
{with outliers}&{MCKF}&{$ 0.3803\pm0.0524$}&{$0.3655\pm0.0785$}&{$0.1495\pm0.0867$}&{$0.1295\pm0.0192$}\\
{}&{MEE-KF}&{$\textbf{0.2785}\pm\textbf{0.0368}$}&{$ \textbf{0.1794}\pm\textbf{0.0478}$}&{$ \textbf{0.1377}\pm\textbf{0.0467}$}&{$\textbf{0.1155}\pm\textbf{0.0245}$}\\
{{mixture Gaussian}}&{KF}&{$0.4903\pm0.0887$}&{$ {0.4840}\pm{0.0375}$}&{$ 0.1572\pm0.0363$}&{$0.1292\pm0.0889$}\\
{noise}&{MCKF}&{$0.4452\pm0.0005$}&{$0.4376\pm 0.0365$}&{$0.1527\pm0.0770$}&{$0.1348\pm0.0699$}\\
{}&{MEE-KF}&{$\textbf{0.2714}\pm\textbf{0.0314}$}&{$ \textbf{0.1796}\pm\textbf{0.0962}$}&{$\textbf{0.1364}\pm\textbf{0.070}$}&{$ \textbf{0.1161}\pm\textbf{0.0368}$}\\
{{mixture Gaussian}}&{KF}&{$1.1973\pm0.0369$}&{$1.1918\pm0.0468$}&{$ 0.3911\pm0.0126$}&{$0.3640\pm0.0469$}\\
{noise with outliers}&{MCKF}&{$0.8420\pm0.0789$}&{$0.8256\pm0.0546$}&{$0.3604\pm0.0345$}&{$ 0.3330\pm0.0549$}\\
{}&{MEE-KF}&{$\textbf{0.6087}\pm\textbf{0.0524}$}&{$\textbf{0.4998}\pm\textbf{0.0479}$}&{$ \textbf{0.3225}\pm\textbf{0.0425}$}&{$\textbf{0.3161}\pm\textbf{0.0245}$}\\
\cline{1-6}
\end{tabular}
\label{table1}
\end{center}
\end{table*}

\section{Experimental Results}
In this section, the performances of the proposed MEE-KF and MEE-EKF are demonstrated in different scenarios through three examples, i.e., land vehicle navigation, tracking of autonomous driving \cite{Udacity, Udacity1} and prediction of infectious disease epidemics \cite{DEP}. The results are compared with KF and MCKF  in linear model, and EKF and MCEKF in nonlinear model. The performance evaluation index is defined by the mean square error (MSE):
\begin{align}
\text{MSE}=\frac{1}{N} \sum_{k = 1}^{N} {{||\textbf{x}(k) - \hat{\textbf{x}}(k)||_1}},
\end{align}
where $\hat{\textbf{x}}(k)$ is the estimate of $\textbf{x}(k)$, and ${N}$ is the number of samples. In the simulations, the MSE is computed by averaging over $100$ independent Monte Carlo runs.

\subsection{Land Vehicle Navigation}
In the first example, we assume a linear land vehicle navigation problem to illustrate the performance of MEE-KF.
\subsubsection{Model}
The state and measurement equations are described by
\begin{align}
\begin{array}{l}
{\rm{\textbf{x}}}(k) =\left[ {\begin{array}{*{20}{c}}
{\rm{1}}&{\rm{0}}&{\Delta T}&{\rm{0}}\\
{\rm{0}}&{\rm{1}}&{\rm{0}}&{\Delta T}\\
{\rm{0}}&{\rm{0}}&{\rm{1}}&{\rm{0}}\\
{\rm{0}}&{\rm{0}}&{\rm{0}}&{\rm{1}}
\end{array}} \right]{\rm{\textbf{x}(}}k{\rm{ - 1)}}+
\textbf{q}(k - 1)\\
\end{array}
\end{align}
\begin{align}
\begin{array}{l}
\textbf{y}(k) = \left[ {\begin{array}{*{20}{c}}
-1&0&-1&0\\
0&-1&0&-1
\end{array}} \right]{\rm{\textbf{x}(}}k{\rm{ - 1)}} + \textbf{r}(k),
\end{array}
\end{align}
where $\textbf{x}(k)=[\textrm{x}_1(k)\,\,\, \textrm{x}_2(k) \,\,\,\textrm{x}_3(k) \,\,\, \textrm{x}_4(k)]^{\textrm{T}}$ is the state
vector with components being the north position, the east position, the north velocity and the east velocity. The variables $\theta=\pi/3$ and ${\Delta T}=0.3$ are the direction, and time interval of vehicle, respectively. The real state $\textbf{x}(0)$, the prior estimate ${\hat{\textbf{x}}}(1|0)$ and the corresponding error covariance matrix ${{\textbf{P}}}(1|0)$ are initialized by
\begin{align}
&\textbf{x}(0)=[0\,\,\, 0 \,\,\,10\textrm{tan}\theta \,\,\, 10]^{\textrm{T}}\\
& {\hat{\textbf{x}}}(1|0)=[1\,\,\, 1\,\,\, 1 \,\,\, 1]^{\textrm{T}}\\
& {{\textbf{P}}}(1|0)=\textrm{diag}[900\,\,\, 900\,\,\, 4 \,\,\, 4].
\end{align}
In the dynamic system, the process noises are assumed to be of the Gaussian distribution with $\textbf{q}(k)\sim \mathcal{N}(0,0.01)$, where $\mathcal{N}(0,0.01)$ denotes a zero-mean Gaussian distribution with variance $0.01$.
\subsubsection{Estimation results of different algorithms}

\begin{table}[!tb]
\caption{Parameter Settings of MCKF and MEE-KF}
\begin{center}
\begin{tabular}{p{4cm} c c}
\hline
{Algorithms}&MCKF&MEE-KF\\
\cline{1-3}
{Parameters}&{$\sigma$}\,\,\,\,\,\,\,\,{$\varepsilon$}&{$\sigma$}\,\,\,\,\,\,\,\,{$\varepsilon$}\\
\cline{1-3}
{Gaussian noise}&{10}\,\,\,\,\,$15^{-6}$&10 \,\,\,\,\,$10^{-6}$\\
{Gaussian noise with outliers}&{6.0}\,\,\,\,\,$10^{-6}$&2.0 \,\,\,\,\,$10^{-6}$\\
{mixture Gaussian noise}&{6.0}\,\,\,\,\,$10^{-6}$&2.0 \,\,\,\,\,$10^{-6}$\\
{mixture Gaussian noise with outliers}&{5.0}\,\,\,\,\,$10^{-6}$&1.5 \,\,\,\,\,$10^{-6}$\\
\cline{1-3}
\end{tabular}
\label{sencondtable1}
\end{center}
\end{table}
The following simulation results are presented to confirm the desirable performance of the MEE-KF. Here, we use $30000$ samples (time steps) to calculate MSE.

First, the influence of the measurement noises on MSE is investigated. For the distribution of $\textbf{r}(k)$, we consider four cases:
(1) Gaussian noise: $\textbf{r}(k) \sim \mathcal{N}(0,0.05)$; (2) Gaussian noise with outliers: $\textbf{r}(k) \sim 0.99\mathcal{N}(0,0.009)+ 0.01\mathcal{N}(0,1000)$, where $\mathcal{N}(0,1000)$ is used to generate outliers; (3) mixture Gaussian noise: $\textbf{r}(k)\sim 0.01\mathcal{N}(-0.1,0.001)+0.99\mathcal{N}(0.1,1000)$;  (4) mixture Gaussian noise with outliers: $\textbf{r}(k)\sim 0.48\mathcal{N}(-0.1,0.001)+0.04\mathcal{N}(0,1000)+0.48\mathcal{N}(0.1,0.001)$.
\begin{table*}
\caption{Estimation Results of Different Algorithms with Different Kernel Sizes}
\begin{center}
\begin{tabular}{l l l l l l }
\hline
\multirow{1}{*}{Algorithms}&{$\sigma$}&{MSE of x1 }&{MSE of x2 }&{MSE of x3 }&{MSE of x4 }\\
\cline{1-6}
\multirow{1}{*}{KF}&N/A&{$1.2031\pm0.3572$}&{$1.1811\pm0.2139$}&{$0.3901\pm0.6712$}&{$0.3622\pm0.2057$}\\
\multirow{1}{*}{MCKF}&$\sigma=3.0$&{$N/A$}&{$N/A$}&{$N/A$}&{$N/A$}\\
&$\sigma=4.0$&{$0.8213\pm0.5663$}&{$ 0.8437\pm 0.3573$}&{$0.3773\pm0.7391$}&{$0.3298\pm0.2330$}\\
&$\sigma=5.0$&{$0.8423\pm0.4875$}&{$0.8121\pm0.2310$}&{$0.3610\pm0.0495$}&{$0.3260\pm0.0352$}\\
&$\sigma=10.0$&{$1.0576\pm0.3515$}&{$1.0366\pm 0.2103$}&{$0.3766\pm 0.6477$}&{$0.3471\pm0.2066$}\\
&$\sigma=20$&{$1.1668\pm 0.3709$}&{$1.1748\pm0.2570$}&{$0.3895\pm0.6414$}&{$0.3615\pm0.2667$}\\
\multirow{1}{*}{MEE-KF}&$\sigma=1.0$&{$1.7206\pm0.3869$}&{$0.5587\pm0.3001$}&{$0.3593\pm0.5045$}&{$0.2896\pm0.1772$}\\
&{$\sigma=1.5$}&{$\textbf{0.6258}\pm\textbf{0.2942}$}&{$\textbf{0.5376}\pm\textbf{0.4943}$}&{$\textbf{0.3296}\pm\textbf{0.4930}$}&{$\textbf{0.2896}\pm\textbf{0.2441}$}\\
&{$\sigma=2.0$}&{$0.6833\pm0.3411$}&{$0.6671\pm0.6711$}&{$0.3490\pm0.4930$}&{$0.4213\pm0.2453$}\\
&$\sigma=3.0$&{$0.8410\pm0.2367$}&{$0.9269\pm 0.2604$}&{$0.3945\pm0.4748$}&{$ 0.5876\pm0.3474$}\\
&$\sigma=5.0$&{$1.0381\pm0.2719$}&{$1.1381\pm0.3910$}&{$0.4391\pm0.4750$}&{$0.7279\pm0.2161$}\\
&$\sigma=10$&{$1.1912\pm0.2995$}&{$1.2142\pm0.3956$}&{$ 0.4628\pm0.4698$}&{$0.7814\pm0.2204$}\\
\cline{1-6}
\end{tabular}
\label{table2}
\end{center}\textit{}
\end{table*}
Table \ref{table1} gives the estimation results of KF, MCKF and MEE-KF in the presence of different measurement noises.  The parameter settings of MCKF and MEE-KF are given in Table \ref{sencondtable1}, and all these parameters are set by trials to obtain the best performance.  From Table \ref{table1}, we observe: i) under the Gaussian noises, the performance of MCKF approaches that of KF, and KF preforms the optimal performance than MEE-KF; ii) when large outliers or mixture Gaussian
noise  (Case (2), Case (3) and Case (4)) are included in the measurement noise, the MCKF and MEE-KF show better performance than KF, and the MEE-KF achieves the best performance among the three algorithms.

Next, we investigate the influence of kernel size on the estimation performance. Table \ref{table2} shows the MSE comparison of different filters with different kernel sizes, where the measurement noise is the same as Case (4). The thresholds in MCKF and MEE-KF are set to $10^{-6}$.  From Table \ref{table2}, we find out that if the kernel size is too small or too large, the performance of MCKF and MEE-KF will become worse. Especially, for the small kernel size, the MCKF may diverge, while MEE-KF is more stable compared with MCKF. The desirable performance can be achieved by MEE-KF when the kernel size is around $1.5$.

Finally, the computing time of different filters at each iteration is summarized in Table \ref{sencondtable1111}, which is  measured with MATLAB $2016$b running on i$5$-$4590$ and $3.30$ GHZ CPU. We see that the MEE-KF can achieve better performance at the cost of slightly higher computational burden than MCKF.

\subsection{Tracking of Autonomous Driving}
\begin{figure}[!t]
\centering
\includegraphics[width=9cm,height=5cm]{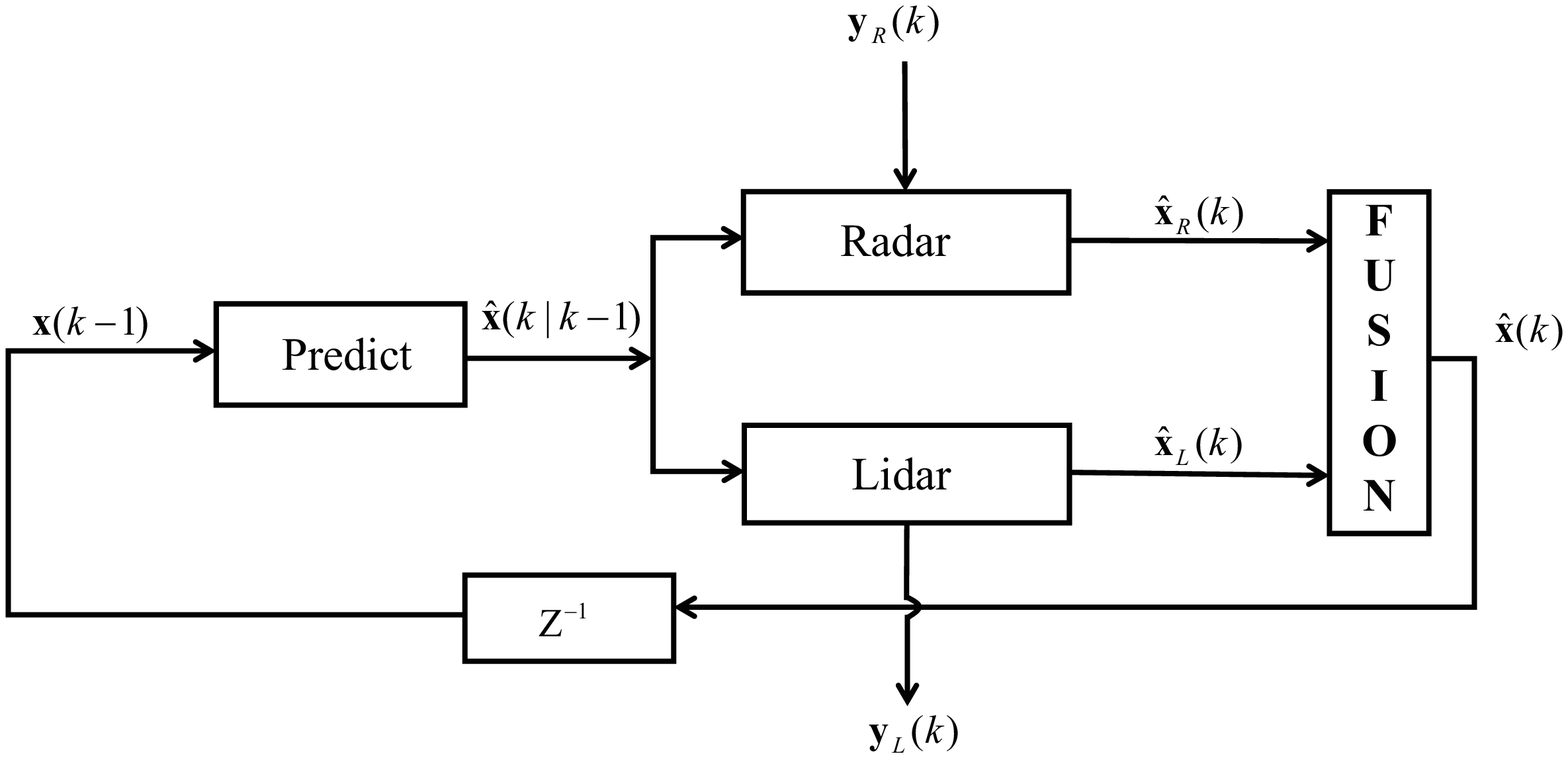}
\caption{Kalman fusion model.}
\label{Figure1}
\end{figure}

This experiment shows the tracking of autonomous driving with the Kalman fusion algorithms \cite{Fusion}. The data source is from Udacity course self-driving car \cite{Udacity}, which is measured by the lidar and radar sensors. In general, different sensors provide different types of information with different accuracies, especially in different weather conditions \cite{Udacity1}. In this example, the lidar sensor provides a good resolution about the position, while the radar sensor provides better accuracy about the velocity in poor weather compared to lidar. To merge the information from different sensors, the Kalman fusion algorithms are applied to different sensors to estimate effectively the trajectory of vehicle.  Fig. \ref{Figure1} shows the Kalman fusion model, where $\textbf{x}(k-1)$ denotes the state variables at time $k-1$; $\textbf{y}_L(k)$ and $\textbf{y}_R(k)$ are the observations from lidar and radar sensors; ${\hat{\textbf{x}}}_L(k)$ and ${\hat{\textbf{x}}}_R(k)$ are the estimated states; ${\hat{\textbf{x}}}(k)$ is the fusion state at time $k$. The detailed descriptions are presented in the following.
\begin{table}[!tb]
\caption{Average Computing Time (sec) for Each Iteration}
\begin{center}
\begin{tabular}{l l l l }
\hline
\multirow{1}{*}{Algorithms}&{KF}&MCKF&MEE-KF\\
\cline{1-4}
\multirow{1}{*}{Computing time(sec)}&{0.000066}&0.000204&0.000232\\
\cline{1-4}
\end{tabular}
\label{sencondtable1111}
\end{center}
\end{table}

\subsubsection{Model}
For the tracked vehicle, the state can be described by a $4$-dimensional vector ${\rm{\textbf{x}}}(k) = [{p_x}(k),{p_y}(k),{v_{x}}(k),{v_y}(k)]^{\textrm{T}}$, where ${p_x}(k)$ and ${p_y}(k)$ denote the position information of $x$ and $y$ coordinates; ${v_x}(k)$ and ${v_y}(k)$ are the corresponding velocity information. The state equation can be expressed by
\begin{align}\label{state92}
\textbf{x}(k) = \left[ {\begin{array}{*{20}{c}}
1&0&{\Delta T}&0\\
0&1&0&{\Delta T}\\
0&0&1&0\\
0&0&0&1
\end{array}} \right]\textbf{x}(k - 1) + \textbf{q}(k),
\end{align}
where ${\Delta T}=0.1$sec is the time interval and $\textbf{q}(k)$ is the process noise with covariance matrix $\textbf{Q}(k) = \left[ {\begin{array}{*{20}{c}}
{\frac{{\Delta {T^2}}}{4}}&0&{\frac{{\Delta {T^3}}}{2}}&0\\
0&{\frac{{\Delta {T^2}}}{4}}&0&{\frac{{\Delta {T^3}}}{2}}\\
{\frac{{\Delta {T^3}}}{2}}&0&{\Delta {T^2}}&0\\
0&{\frac{{\Delta {T^3}}}{2}}&0&{\Delta {T^2}}
\end{array}} \right]$. The prior error covariance matrix is initialized as
${\textbf{P}}(1|0) = \textrm{diag}\left[1,1,1000,1000 \right]$.

\begin{figure}[!t]
\centering
\includegraphics[width=7cm,height=5.5cm]{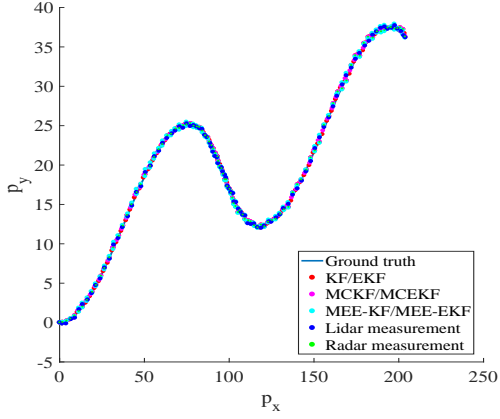}
\caption{Tracking results of different filters for trajectory 1.}
\label{Figure2}
\end{figure}
For the measurement equation, we consider two cases:\\
\emph{Case I}: When the measurements $\textbf{y}_L(k)=[{p_x}(k),{p_y}(k)]^{\textrm{T}}$ are acquired from the lidar sensor, the measurement equation can be described as
\begin{align}\label{measurement93}
\textbf{y}_L(k) = \left[ {\begin{array}{*{20}{c}}
1&0&0&0\\
0&1&0&0
\end{array}} \right]\textbf{x}(k) + \textbf{r}_L(k),
\end{align}
where $\textbf{r}_L(k)$ is the measurement noise from lidar sensor with  covariance matrix ${\textbf{R}_L}(k) = \textrm{diag}\left[ 0.0025,0.0025
 \right]$. For the linear state space model in (\ref{state92}) and (\ref{measurement93}), the KF, MCKF and MEE-KF can be employed. \\
\begin{figure}[!t]
\centering
\includegraphics[width=7cm,height=5.5cm]{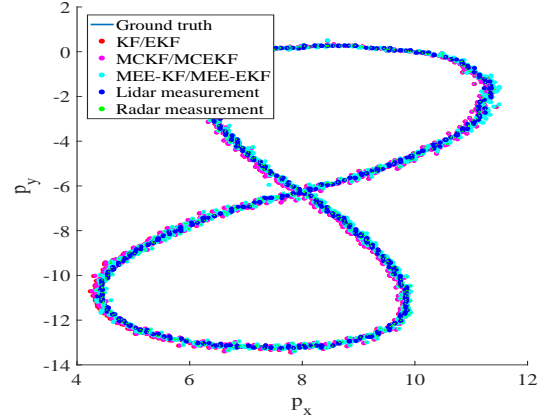}
\caption{Tracking results of different filters for trajectory 2.}
\label{Figure3}
\end{figure}
\begin{table*}[!t]
\caption{Estimation Results of Different Algorithms for Trajectory $1$ and Trajectory $2$}
\begin{center}
\begin{tabular}{l l l l}
\hline
\multirow{1}{*}{Algorithms}&{KF/EKF}&MCKF/MCEKF&MEE-KF/MEE-EKF\\
\cline{1-4}
\multirow{1}{*}{MSE of trajectory 1}&{0.5408}&0.2570& 0.1554\\
\multirow{1}{*}{MSE of trajectory 2}&{0.008887}&0.008184& 0.007789\\
\cline{1-4}
\end{tabular}
\label{sencondtable}
\end{center}
\end{table*}
\emph{Case II}: The measurements $\textbf{y}_R(k)=\left[ {\begin{array}{*{20}{c}}
{\rho (k)},{\varphi (k)},{\bar \rho (k)}
\end{array}} \right]^{\text{T}}$ from the radar sensor involve three components in polar coordinates, i.e., the range $\rho(k)$, the angle $\varphi (k)$ between $\rho(k)$ and $x$ coordinates axis, and the range rate ${\bar \rho (k)}$.
The measurement equation can then be established by:
\begin{align}\label{measurement95}
\textbf{y}_R(k) = \textbf{h}(\textbf{x}(k)) + \textbf{r}_R(k),
\end{align}
where the nonlinear function
\begin{align}
\textbf{h}(\textbf{x}(k))=\left[ {\begin{array}{*{20}{c}}
{\sqrt {p{}_{_{{x}}}^{{2}}(k){{ + }}p{}_{_{{y}}}^{{2}}(k)} }\\
{\arctan (\dfrac{{{{{p}}_y}(k)}}{{{{{p}}_x}(k)}})}\\
{\dfrac{{{{{p}}_x}(k){{{v}}_x}(k) + {{{p}}_y}(k){{{v}}_y}(k)}}{{\sqrt {p{}_{_{{x}}}^{{2}}(k){\rm{ + }}p{}_{_{{y}}}^{{2}}(k)} }}}
\end{array}} \right]
\end{align}
can be derived by a mapping from the polar coordinates to the cartesian coordinates; $\textbf{r}_R(k)$ is the measurement noise from radar sensor with covariance matrix ${\textbf{R}_R}(k) = \textrm{diag} \left[ 0.09,0.05,0.09\right]$. To estimate the state $\textbf{x}(k)$ in the cartesian coordinates, one should transform the measurements $\textbf{y}_R(k)=\left[ {\begin{array}{*{20}{c}}
{\rho (k)},{\varphi (k)},{\bar \rho (k)}
\end{array}} \right]^{\text{T}}$ as state vector $ \textbf{x}(k)=\left[{p_x}(k),{p_y}(k),{v_{x}}(k),{v_y}(k)\right]^{\text{T}}$:
\begin{align}
\begin{array}{l}
{p_x}(k) = \rho (k)\cos (\varphi (k)),\,\,\,\,\,\,{p_y}(k) = \rho (k)\sin (\varphi (k))\\
{v_{x}}(k) = \bar \rho (k)\cos (\varphi (k)),\,\,\,\,\,\,{v_y}(k) = \bar \rho (k)\sin (\varphi (k)).
\end{array}
\end{align}

For the nonlinear state space model in (\ref{state92}) and (\ref{measurement95}), the EKF, MCEKF and MEE-EKF can be used.

\subsubsection{Estimation results of different algorithms}
The compared algorithms are performed for two trajectories, i.e., trajectory $1$ and trajectory $2$. Fig. \ref{Figure2} and Fig. \ref{Figure3} show the tracking results of trajectory $1$ and trajectory $2$, respectively. Table \ref{sencondtable} presents the MSE of different Kalman fusion algorithms, i.e., the KF/EKF, MCKF/MCEKF and  MEE-KF/MEE-EKF. In this simulation, the kernel sizes are set to $1.66$ in MEE-EKF, $20.0$ in MEE-KF, $15.0$ in MCC-EKF and $20.0$ in MCC-KF. From Table \ref{sencondtable}, one can see that the MEE-KF/MEE-EKF fusion method exhibits the best performance among all compared algorithms.

\subsection{Prediction of Infectious Disease Epidemics}
In this part, the prediction of epidemic \cite{DEP} is used to validate the effectiveness of the proposed MEE-EKF. The disease control center (DCC) provides the Influenza-Like-Illness (ILI) data \cite{DEP} of the United States, a traditional information source for the epidemic estimation.
\subsubsection{Model}
The state process of an epidemic is based on the Susceptible-Infected-Recovered (SIR) model~\cite{DEP}, given by
\begin{align}
&s(k + 1) = s(k)- \delta_1 s(k)i(k)\label{93}\\
&i(k + 1) = i(k) + \delta_1 s(k)i(k) - \delta_2 i(k)\label{94}\\
&b(k + 1) = b(k) + \delta_2 i(k),
\end{align}
where $s(k)$, $i(k)$ and $b(k)$ are the susceptible,  infected and recovered crowd density, respectively, with $s(k)+i(k)+b(k)=1$ for all $k$.
The parameter $\delta_1$ denotes the rate about the spread of illness and $\delta_2$ is the rate about recovery from infection. Since the recovered $b(k)$ is obtained by $b(k) = 1-s(k)-i(k)$, the state of epidemic at time $k$ can be described by $\textbf{x}(k) = [s(k), i(k)]^{\textrm{T}}$. Therefore, the state equation can be written by
\begin{align}
\textbf{x}(k) = \textbf{f}(\textbf{x}(k-1)) + \textbf{q}(k),
\end{align}
where the nonlinear function $\textbf{f}(\textbf{x}(k-1))$ is given in Eqs. (\ref{93}) and (\ref{94}),  and $\textbf{Q}(k)=10^{-8}\textbf{I}_n$.

\begin{figure}
\centering
\includegraphics[width=7cm,height=5.5cm]{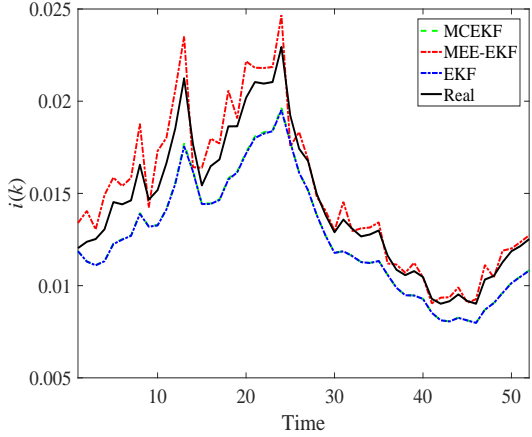}
\caption{Estimation results for infected $i(k)$ by EKF, MCEKF and MEE-EKF.}
\label{Figure4}
\end{figure}

In the state space model, the measurement equation specifies how the observed data depend on the state of epidemic. When monitoring an epidemic, the unknown $s(k)$, $i(k)$, and $b(k)$ are regarded as hidden states of the model, and the observed data are acquired via syndromic surveillance system. Due to only $i(k)$ can be obtained by the syndromic surveillance system, the measurement equation can be established by
\begin{align}
{y}(k) = \textrm{\textbf{H}}\textbf{x}(k) + {r}(k),
\end{align}
where the matrix $\textbf{H}=[0\,\,\, 1]$ denotes the measurement matrix and  $\textbf{R}(k)=10^{-6}$.
\begin{table}[!t]
\caption{Performance Comparison of Different Algorithms for Prediction of Infectious Disease Epidemics}
\begin{center}
\begin{tabular}{l l l l}
\hline
\multirow{1}{*}{Algorithms}&{EKF}&MCEKF&MEE-EKF\\
\cline{1-4}
\multirow{1}{*}{MSE}&{0.001729}&0.001703& 0.000825\\
\cline{1-4}
\end{tabular}
\label{tab:sencondtable11}
\end{center}
\end{table}
\subsubsection{Estimation results of different algorithms}
Figure \ref{Figure4} plots $52$ weeks density of people infected in $2010$ year.  Table \ref{tab:sencondtable11} gives the estimated MSE  by EKF, MCEKF and MEE-EKF algorithms.  In Fig. \ref{Figure4}, a peak around the $25th$ week occurs, which means that the number people infected  arrives at maximum. In the SIR model, the contact rate $\delta_1$ and recovery time $\delta_2$ are set to $0.35$ and $0.11$, respectively. The kernel sizes of MEE-EKF and MCEKF are $20.0$, and $16.0$, respectively. From Fig. \ref{Figure4} and Table \ref{tab:sencondtable11}, the MEE-EKF provides a much better prediction for the density of people infected than EKF and MCEKF.

\section{Conclusion}
In this paper, we propose the minimum error entropy Kalman filter (MEE-KF) with a fixed-point iteration. Unlike the original Kalman filter (KF) based on the well-known minimum mean square error (MMSE) criterion and the maximum correntropy Kalman filter (MCKF) based on the maximum correntropy criterion (MCC), the MEE-KF is developed by using the minimum error entropy (MEE) criterion as the optimality criterion.  With an appropriate kernel size, the MEE-KF algorithm can achieve better performance than the KF and MCKF  especially when the underlying system is disturbed by some complicated non-Gaussian noises. The computational complexity of the MEE-KF is provided, and a sufficient condition for the convergence of the fixed-point iteration is given. Furthermore, the MEE based extended Kalman filter (MEE-EKF) is also developed to deal with the problem of state estimation of a nonlinear system in non-Gaussian noises. Simulations on land vehicle navigation, tracking of autonomous driving and prediction of infectious disease epidemics have confirmed the desirable performance of the proposed MEE-KF and MEE-EKF.

\begin{ack}
This work was supported by the 973 Program (No. 2015CB351703),  National Key R\&D Program of China (No. 2017YFB1002501), and  National NSF of China (No. 91648208, No. U1613219).
\end{ack}

\end{document}